%% file: IEEE-conference.tex
\documentclass[conference]{IEEEtran}
\IEEEoverridecommandlockouts

\usepackage{cite}
\usepackage{amsmath,amssymb,amsfonts}
\usepackage{algorithmic}
\usepackage{graphicx}
\usepackage{textcomp}
\usepackage{xcolor}
\usepackage{subfig}
\def\BibTeX{{\rm B\kern-.05em{\sc i\kern-.025em b}\kern-.08em
    T\kern-.1667em\lower.7ex\hbox{E}\kern-.125emX}}

\begin{document}

\title{ RIS-Aided Radar Imaging Utilizing the Virtual Source Principle
\thanks{This work is funded by the German Research Foundation (“Deutsche Forschungsgemeinschaft”) (DFG) under Project–ID 287022738 TRR 196 for Project S03 and Project–ID GZ: SE 1697/21-1.} }

\author{\IEEEauthorblockN{Furkan H. Ilgac}
\IEEEauthorblockA{\textit{Ruhr University Bochum} \\
Bochum, Germany \\
Furkan.Ilgac@ruhr-uni-bochum.de}
\and
\IEEEauthorblockN{Mounia Bouabdellah}
\IEEEauthorblockA{\textit{Ruhr University Bochum} \\
Bochum, Germany \\
Mounia.Bouabdellah@ruhr-uni-bochum.de}
\and
\IEEEauthorblockN{Aydin Sezgin}
\IEEEauthorblockA{\textit{Ruhr University Bochum} \\
Bochum, Germany \\
Aydin.Sezgin@ruhr-uni-bochum.de}}

\maketitle

\begin{abstract}
This paper investigates signle-antenna radar imaging with a reconfigurable intelligent surface (RIS). Configuring phase shifts in a RIS-aided radar system can be thought as synthetic aperture radar (SAR) imaging with a moving virtual source. With this perspective, the problem is modeled in the wavenumber domain and image forming algorithms are formulated for near and far field regions.
\end{abstract}


\section{Introduction}
\input{introduction.tex}
\section{System Model}
\input{system_model.tex}

\subsection{Near Field Formulation}
The system geometry for the nearfield is illustrated in Fig. 3. In this model, the wavefronts travel as spherical waves rather than the planar ones, therefore, the received signal is in the general form 
\begin{equation} \label{eq:yioNearfield}
y(i,\phi_n) = \sigma_p A 
\sum_{n=0}^{N-1}
\exp \left[
-j k_i({r}_n+s_n+\phi_n+
p )
\right].
\end{equation}
The geometrical relationships between the different paths lead to highly non-linear equations. As a consequence, received signal can not be simplified as in the case of eq. (\ref{eq:y_io}). Nonetheless, the observation we have made in the previous subsection can be extended here, which indicates us that the azimuth term in eq. (\ref{eq:yioNearfield}) correspond to takin non-uniform Fourier transformation (NUFT) of the phase delays in RIS elements with respect to non-uniformly placed path differences. Consequently, we might apply windows specialized for  non-uniform transformation to enhance the characteristics of the resulting image \cite{nfft}. However, in this case, the sampling points that we need to apply the window is dependent on the problem geometry, and consequently changes over the scene of interest based on the (prospecitve) targets' location, which can be formulated by
\begin{equation} \label{eq:yioNearfieldWindow}
y(i,\phi_n) = \sigma_p A 
\sum_{n=0}^{N-1}
w(n,p)
\exp \left[
-j k_i({r}_n+s_n+\phi_n+
p )
\right].
\end{equation}
Practically, to create an image under these conditions, we might divide the scene into sub-regions, create windowing functions for the corresponding sub-region centers, and perform backprojection by inverting the signal, i.e. multiplying the eq. (\ref{eq:yioNearfieldWindow}) with its approximate matched filter. An appropriate choice for improving side-lobe levels in non-uniform Fourier Transformations could be the Gaussian filter, which is given by
\begin{equation}
    w(n,p) = \frac{1}{\sqrt{2\pi\sigma_l}} 
    \exp{\left[
        \frac{
        (l_{n,p}-\mu_l)^2
        }{
        \sigma_l^2
        }\right]
        }
\end{equation}
where $l_{n,p} = s_n+r_n+p$ is the axillary variable denoting the sampling points in the NUFT, $\mu_l$ and $\sigma_l$ are the mean and the sample variance of the sample points, respectively. 

\begin{figure} [ht]
    \centering
    \includegraphics[width=0.75\linewidth]{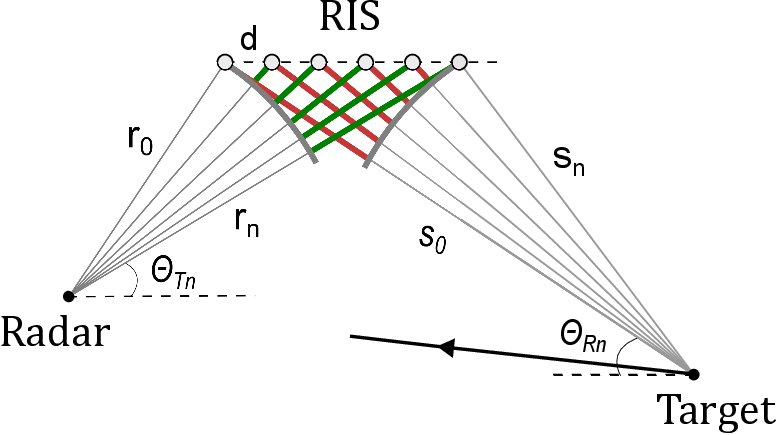}
    \caption{Near field geometry}
    \label{fig:nearfield_model}
\end{figure}

\section{Results}
\begin{figure*}[h]
    \centering
  \subfloat[ SAR imaging with ground truth.]{
        \includegraphics[scale=0.40]{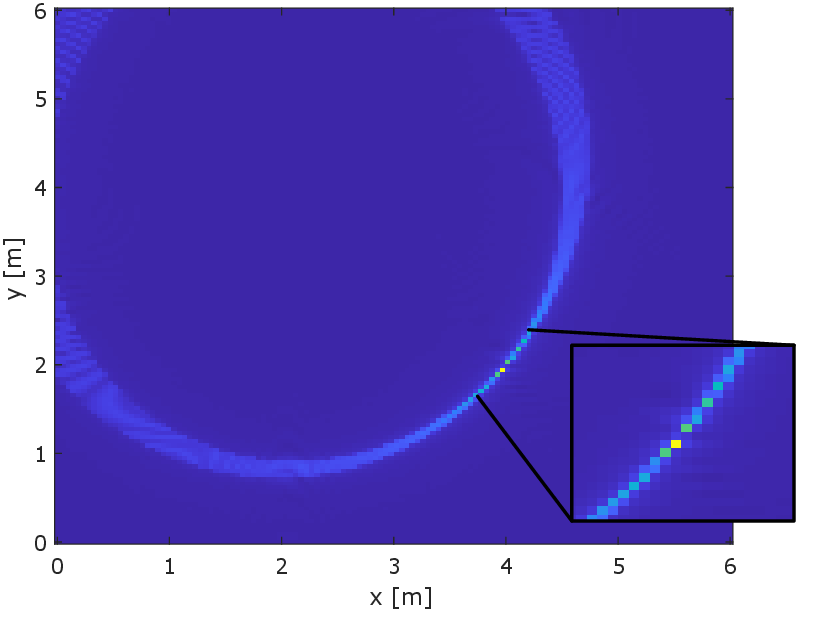}
    }
  \subfloat[Virtual source imaging. \label{fig:dif 0dB}]{
        \includegraphics[scale= 0.40]{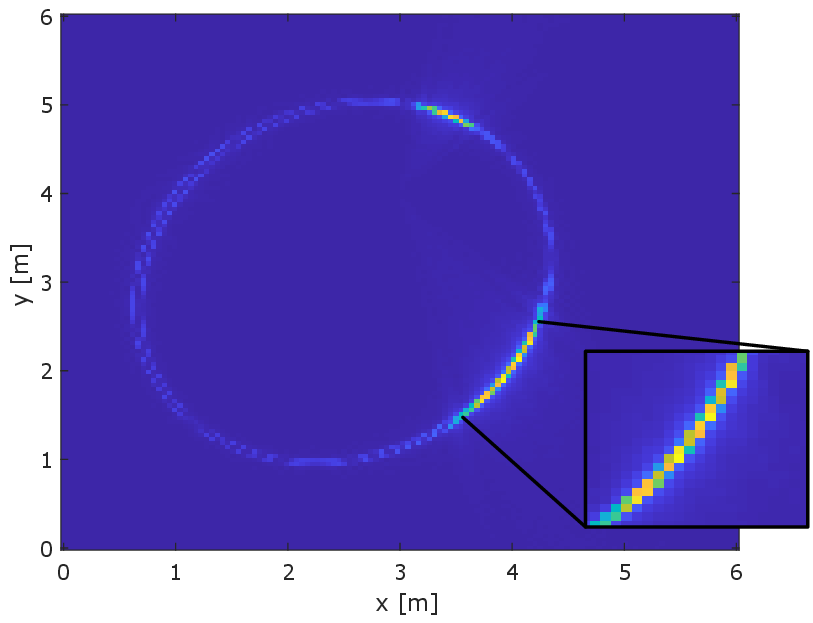}
    }
  \subfloat[ Virtual source imaging with Hamming window.\label{fig:dif40dB }]{
       \includegraphics[scale= 0.40]{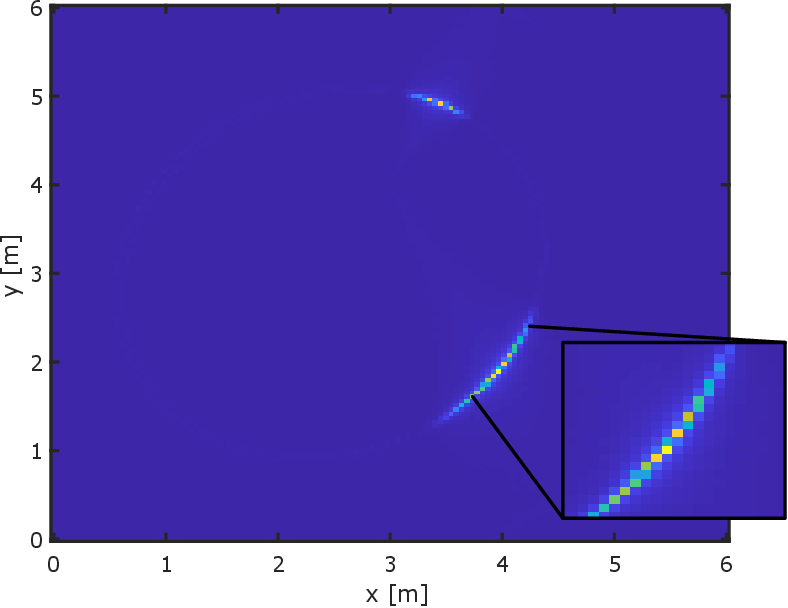}
   }
   \caption{ Far field imaging results. }
    \end{figure*}
    \begin{figure} [ht]
    \centering
    \includegraphics[width= 0.95
    \linewidth]{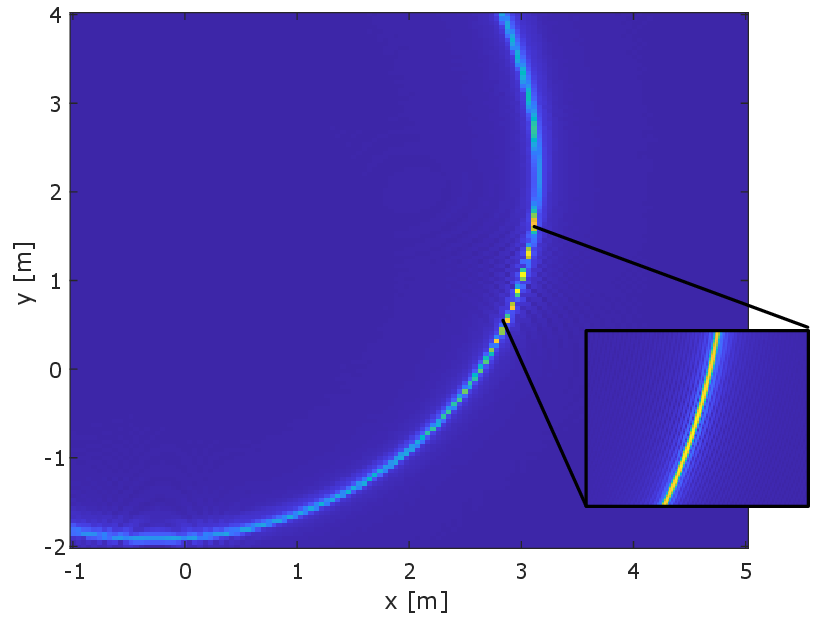}
    \caption{Near field image formation without any windowing.}
    \label{fig:nearfield_model}
\end{figure}
\begin{figure} [ht]
    \centering
    \includegraphics[width= 0.95
    \linewidth]{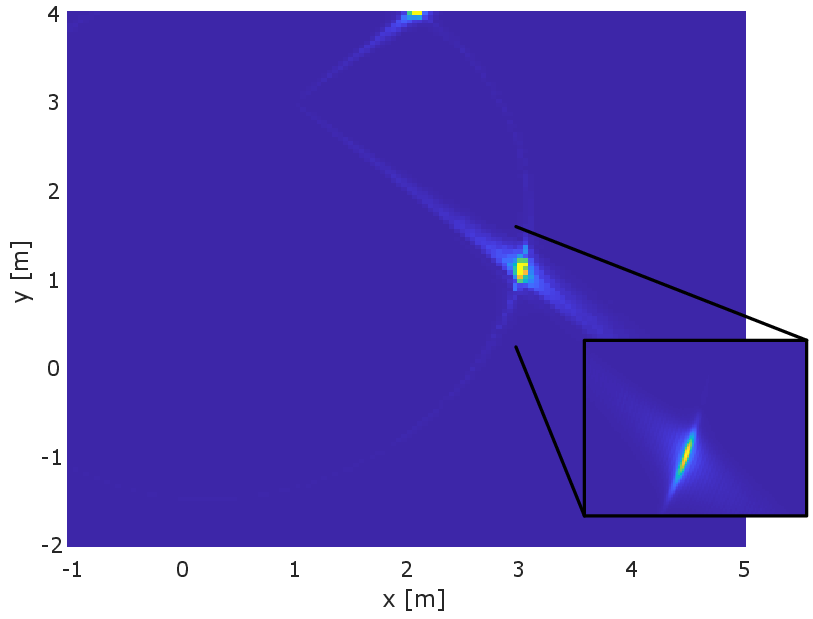}
    \caption{Near field image formation with Gaussian window.}
    \label{fig:nearfield_model}
\end{figure}
In this section, we validate the studies conducted for the proposed radar imaging method. We model an indoor FMCW radar operating at 120 GHz with a 10 GHz bandwidth, in conjunction with a reflective-type RIS. The FMCW signal has a period of 1$\mu$s and 1024 chirp bits. The RIS-aided radar setup is placed in both near- and far-field configurations, and a single-point-target response is produced. The RIS phases are configured to correspond to a 20-degree movement of the virtual source, divided into 20 equal positions on a circle, where the radius is equal to the distance between the radar and the RIS center. The simulations are performed with 20dB singal-to-noise ratio (SNR) at the receiver.

In the far field experiments, the locations of the radar, RIS and the target are set as $(2,2),(3,4),(4,2)$, respectively. RIS is assumed to be an ULA with 16 elements, corresponding to a far-field distance of 0.28 meters. In Fig. 4, a comparison between the SAR image and the proposed RIS-aided virtual source imaging method is presented. It is observed that the SAR image provides the highest quality target response which is expected, as the wave field impinging on the target is unrestricted. In contrast, the RIS-aided radar virtual source imaging demonstrates the side lobes due to the rectangular windowing effect discussed in previous sections, which can be seen as repetitive echoes along the azimuth axis, as shown in the Fig. 4.b. Nonetheless, these artifacts can be compensated by windowing, such as shown in the Fig. 4.c, where Hamming window is applied to RIS elements along with phase shifts. It can be observed that the azimuth echoes are much more focused in addition to reduced noise level due to this focusing. 
In RAR imaging, the azimuth resolution is determined by the formula $\delta x_{\text{RAR}} = R \lambda/ D$, where  $R$ is the distance between radar and the target, $\lambda$ is the wavelength and  $D$ is the aperture size.  Similarly, for SAR imaging, the azimuth resolution is given by  $\delta x_{\text{SAR}} = \lambda/ 2L $, where $L$ is the length of the synthetic aperture \cite{Curlander}. To compare, in a system using an aperture equivalent to the size of  the RIS given in this experiment, the RAR resolution is calculated as  $\delta x_{\text{RAR}} = 0.2795$m. Similarly, the theoretical azimuth resolution in  SAR approach equivalent to motion of the virtual source is much finer, $\delta x_{\text{SAR}} = 0.0032$m. Although, in practice, the motion errors that can significantly degrade image quality, particularly at such small wavelengths. The empirical resolution achieved by the proposed windowed method, as shown in Fig. 4.c, is approximately $0.1000$ m. This result demonstrates a satisfactory compromise between the two baselines, achieving better a performance than the RAR, without the need for mechanical movement of the radar platform.

  In near field experiment, the locations of the radar, RIS and the target are set as $(0,0),(1,3),(3,1)$, respectively. The RIS assumed to have 128 elements, corresponding to a far-field distance of 20.16 meters. Figure 5 illustrates the near field scenario where no windowing is applied. Again, the resulting image demonstrates that the non-windowed method fails to produce adequate target response. Conversely,  Figure 6 presents the results for the same geometry, but with Gaussian windowing. At each virtual position, a corresponding window is generated and applied for the point symmetrical to virtual source with respect to the RIS center. The comparison highlights the performance improvement with the windowed method, overcoming the limitations of the non-windowed approach. In addition, here we observe a trade-off introduced by the applying the windowing. As it is well known, windowing improves the side lobe level at the expense of increasing the main beam width, which is evident in Figure 6 from the thickened target echo.


\section{Conclusions}
In this study, we have presented a novel perspective on  radar imaging with RIS, based on virtual sources. We have demonstrated that the RIS-assisted SISO radar setup can be effectively configured as a virtual source imaging system by designing appropriate phase shifts in the RIS elements. We have shown that finite boundaries of the RIS introduce a windowing effect to the motion of this virtual source, and by designing and applying correct windowing techniques, these effect can be compensated, resulting in high-quality target echos with the backprojection algorithm.

\bibliographystyle{IEEEtran}
\bibliography{references}

\end{document}

%% file: introduction.tex
The rapid increase in connected devices and data-intensive applications led to an unprecedented demand for ultra-fast, low-latency communication systems. These requirements are further amplified by the vision for 6G networks, which aim to deliver significantly enhanced connectivity. Achieving these goals, however, requires breakthroughs in spectrum efficiency, energy consumption, and signal reliability that conventional wireless infrastructures struggle to provide \cite{tataria20216g}. In this context, reconfigurable intelligent surfaces (RIS) have gained significant attention due to their remarkable flexibility in dynamically shaping the electromagnetic spectrum.

Consisting of planar microstructures with controllable impedances, RIS is a form of metasurface that can alter the phase and amplitude of impinging electromagnetic waves with minimal power consumption. RIS can operate in reflective, transmissive, or hybrid modes \cite{zeng2021reconfigurable}. Reflective RIS bounces signals back toward users on the same side of the base station, while transmissive RIS let signals pass through to serve users on the opposite side. Hybrid RIS can do both, reflecting and transmitting signals as needed. By enabling intelligent reflection and redirection of signals, RIS can create virtual line-of-sight (LoS) paths where direct paths are obstructed, enhancing network reliability and quality of service. Given these capabilities, RIS is not only a powerful tool for communication technology but also an effective resource for wireless sensing and radar imaging.

Wireless sensing represents a significant step forward in communication and environmental monitoring. By analyzing radio frequency signals, these systems can detect and interpret environmental changes, enabling applications like human motion detection, localization, monitoring, etc. By observing how signals interact with the physical surroundings, wireless sensing can identify obstacles, estimate distances, and recognize gestures. Its flexibility makes wireless sensing useful across a variety of fields, from smart home automation to industrial process management and healthcare, providing real-time insights into spatial and situational awareness.

Radar imaging, on the other hand, constitutes a foundational functionality in wireless sensing. Rather than examining characteristics of a single detection at a time, radar imaging aims to extract detailed, picture-like maps of the environment, enabling a wide range of applications, including terrain mapping, environmental monitoring, and advanced surveillance capabilities. In the public domain, one can distinguish between two foundational methods in radar imaging: Real Aperture Radar (RAR) and Synthetic Aperture Radar (SAR). RAR scans a narrow beam across a swath and creates images based on return time. In SAR imaging, the radar moves over a target area to create a synthetic aperture, thus achieving high-resolution imaging. Collecting and combining data from different positions enables SAR to attain high spatial resolution. 

Traditional radar systems rely on the reflection and scattering of electromagnetic waves inevitably bound to work under LOS conditions. Combined with high signal attenuation and complex interference characteristics in next-generation wireless bands, such as the terahertz band, the integration of RIS into radar systems presents a promising solution to improve reliability and performance in future wireless applications. To this end, the RIS has been used to improve the signal-to-noise ratio (SNR), consequently improving target detection performance \cite{buzzi}, enabling sensing of the blocked areas \cite{emrah,ilgac}, and reducing interference \cite{risInterference}. RIS has also been used an extension tool in radar imaging problems. Recent research on RIS-aided imaging has focused on CS-based methods, which adjust RIS phase shifts, capture channel responses, and use these to iteratively generate region of interest (ROI) images, capitalizing on their sparsity in three-dimensional space. For instance, the MetaSketch system \cite{ZhuHan3} applies CS to RF-sensing data, enabling semantic segmentation and object recognition through RIS in complex environments. The MetaSketch reconstructs spatial point clouds from RF signals, which are then semantically segmented using machine learning. FT-based algorithms break the imaging process into one-dimensional operations, which can be efficiently executed using fast Fourier. For example, a FT-based wavenumber domain 3D imaging approach was proposed in \cite{RIS_k_domain}. The technique used focuses on a two-stage process: recovering the equivalent channel response via modified RIS phase shifts, followed by conventional FT-based imaging in the wavenumber domain. This method is particularly advantageous in RIS-aided systems, as it accelerates imaging while managing memory and processing demands effectively. Further advancing radar imaging capabilities, deep learning models have also been incorporated to optimize RIS configurations for 3D RF sensing, as demonstrated in the MetaSensing framework \cite{ZhuHan2}. Using deep reinforcement learning, MetaSensing enhances sensing accuracy by dynamically configuring the RIS for improved object localization. Another approach, coded aperture radar imaging, leverages RIS with variable phase profiles for each radar pulse, using sparse recovery to achieve high-resolution imaging with simplified hardware requirements \cite{codedAperture}. Expanding SAR functionality, range-Doppler (RD) imaging has been integrated with active RIS on mobile platforms (UAV), to create a larger virtual aperture, allowing stationary radar systems to conduct high-resolution imaging of distant targets \cite{RISAidedSAR}. Finally, MetaPhys \cite{li2023metaphys} illustrates the potential of RIS in physiological monitoring, utilizing 4D radar imaging to capture respiratory and cardiac signals of multiple individuals. This method enhances signal-to-noise ratios by beamforming the radiated energy, thus supporting precise physiological sensing in non-line-of-sight (NLOS) environments.

While these advancements in radar imaging techniques, including FT-based, CS-based, deep learning-enhanced methods, etc. have brought improvements in precision and computational efficiency, these approaches introduce new challenges that limit their practicality in certain contexts. For instance, FT-based imaging, while reducing some computational demands, often requires high pilot overhead, which can hinder performance in resource-constrained environments or real-time applications. CS-based methods, are computationally intensive and require increased memory usage, which may restrict their application in low-power or limited-capacity systems. Deep reinforcement learning models could bring higher accuracy in localization but at the cost of significant computational complexity and extensive training times. Although coded aperture radar imaging mitigates hardware needs, it remains sensitive to sparse recovery algorithms, which can introduce trade-offs in image quality under variable environmental conditions. These limitations highlight the need for a simpler approach. Leveraging RIS to create a virtual moving source could address these limitations by simulating movement through phase control rather than physical mobility or computationally intensive algorithms, enabling more adaptable, energy-efficient high-resolution imaging, especially suited to real-time, resource-constrained applications.

In this work, we aim to introduce a novel perspective on RIS-aided radar imaging by invoking image-source concept in the wavenumber domain, and introducing windowing methods for enhancing azimuth compression performance. With this perspective, we aim to extract a bistatic synthetic aperture radar performance from a static SISO RIS-aided radar by synthesizing a virtual source over the real RIS aperture. The rest of this paper is organized as follows, in Section II, we discuss the system model for radar imaging with RIS-aided frequency modulated continous wave (FMCW) radar in different antenna field regions. We discuss the relation between the problem geometry and the received signals, and formulate solutions for near field, reactive near field and far field. Later, in Section III, we provide the results of our methods and compare it with radar images generated by physically moving synthetic aperture radar. Finally, we provide our remarks and conclusions in Section IV.

%% file: system_model.tex
Any signal path resulting from a specular reflection can be modeled by an image-source that mimics the behavior of the actual source, positioned symmetrically across the reflective surface, effectively turning the boundary into an extension of the medium for  analysis. This method has extensively been used in antenna design \cite{balanis} and acoustic localization \cite{acoustics}. Thanks to RIS, we would like to effectively control this image-source, essentially creating a virtual source that can be moved electronically. An illustration of the system geometry is given in Fig. 1, which is essentially a RIS-aided monostatic SISO radar setup. We assume the radar uses FMCW signals to sense the environment, and can control the RIS effectively. 

The FMCW signal transmitted from the radar is given as
\begin{equation} \label{eq:srf}
s_{\text{RF}}(t) = A \exp \left( j 2\pi f_c t + j \pi \gamma t^2 \right),
\end{equation}
where $A$ is the amplitude of the signal, $f_c$ is the carrier frequency, $\gamma$ is the chirp rate, which is obtained by dividing the bandwidth $\beta$ to signal period $T$, and $t$ is fast-time. This signal arrives at a point located at a distance $R_o$ from the radar, with a corresponding time delay $t_o$, resulting in:
\begin{equation}
s_{\text{RF}}(t_o) = A \exp \left( j 2\pi f_c (t-t_o) + j \pi \gamma (t-t_o)^2 \right),
\end{equation}
equivalently
\begin{equation}
s_{\text{RF}}(t_o) = 
A 
s_{\text{RF}}(t) \exp \left( -j 2\pi f_c t_o - j2 \pi \gamma t_o t + j\pi \gamma t_o^2 \right).
\end{equation}
This signal can be converted to baseband by multiplying by the complex conjugate of the original signal characterized by eq. (\ref{eq:srf}). Moreover, the third term in the exponential is called residual video phase (RVP), and changes slowly compared to the first two terms \cite{richards}. Ignoring the RVP, baseband converted signal is,
\begin{equation}
s(t_o) = A \exp \left( -j (2\pi f_c + j2 \pi \gamma t) t_o \right).
\end{equation}
defining the wavenumber $k_i = (2\pi f_c + 2\pi \gamma t)/c$, the equation can be rearranged as
\begin{equation}
s(R_o) =  
A \exp \left( -j k_i R_o\right).
\end{equation}
In the following subsections, we will discuss the formulation for both far-field and near-field scenarios. Since the geometrical assumptions in each zone result in different received signal models, corresponding backprojection algorithms require slightly different modifications or simplifications.

\subsection{Far Field Formulation}
A detailed illustration of the system is given in Fig. 2, where the geometric relations between the radar, target and the two consecutive RIS elements are given. The electromagnetic wavefront is assumed to be traveling as a planar surface. The distance between the consecutive RIS elements are denoted by $d$. In this setup, the radar signals first initially impinge on the RIS, get reflected over with custom phase delays, arrive on target, and finally reflect back to the radar. Assuming no additional phase shift applied, the signal arriving into radar over the RIS antenna $n$ can be written as
\begin{equation} \label{eq:yn}
y_n^{\text{Rx}} = \sigma_p A \exp \left[ -j k_i(|\overline{r}_n| + |\overline{s}_n| + |\overline{p}|)\right].
\end{equation}
\begin{figure} [h]
    \centering
    \includegraphics[width=.8\linewidth]{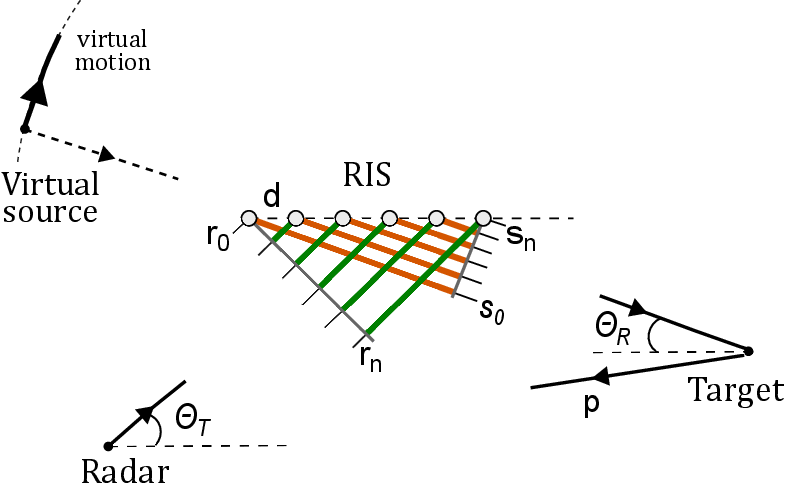}
    \caption{RIS-aided radar imaging with a source image in the far field.}
    \label{fig:far field}
\end{figure}
where $\overline{r}_n$, $\overline{s}_n$, and $\overline{p}$  denote the vectors from the radar to RIS element $n$, RIS element $n$ to target, and target to radar receiver, respectively. By examining the geometric relations between these distances and the angles illustrated in the Fig. 2, we can formulate path differences among different RIS elements. Assuming an imaginary axis that is parallel to the RIS and denote the angles $\theta_T$ and $\theta_R$  correspond to incoming outgoing angles, respectively. Due to far field assumption, the paths could be approximately same with only line segments denoted green and red causing the difference. In that case, the following relations could be formed:
\begin{figure} [ht]
    \centering
    \includegraphics[width=0.7  \linewidth]{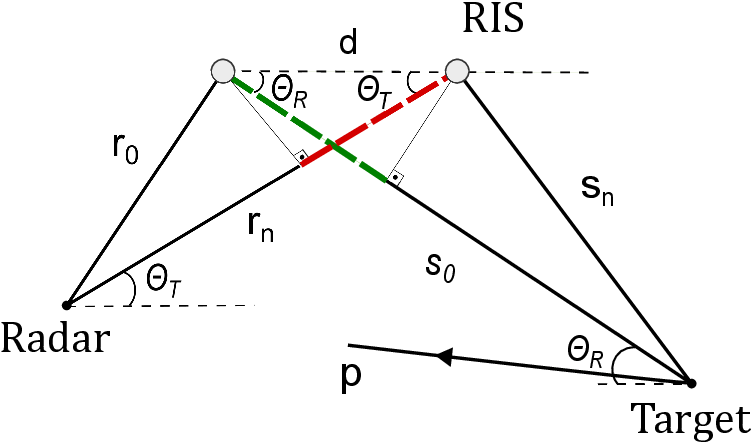}
    \caption{Far field geometry}
    \label{fig:enter-label}
\end{figure}
\begin{equation} \label{eq:FarFieldVectors}
\begin{split}
\overline{r}_n &= |\overline{r}_0| + \overline{r}_0 \cdot \overline{x} 
= r_0 + n d \cos{\theta_T}, \\
\overline{s}_n &= |\overline{s}_0| - \overline{s}_0 \cdot \overline{x}
= s_0 - n d \cos{\theta_R} ,\\
p   &= |\overline{p}| = |r_0 - s_0|.
\end{split}    
\end{equation}
Assuming, the signal returns over the point-target with reflectivity $\sigma_p$, the eq. (\ref{eq:yn}) can be re-written as,
\begin{equation}
\begin{split}
y_n^{\text{Rx}} = 
\sigma_p A 
\exp [ &
-j k_i(r_0 
+ s_0 +  p\\
&+ n d(\cos{\theta_T}-\cos{\theta_R})
].
\end{split}
\end{equation}
more simply,
\begin{equation} \label{eq:y_nVectors}
y_n^{\text{Rx}} = 
\sigma_p A 
\exp \left[-j k_i (r_0 + s_0 + nd\overline{x} \cdot 
(\overline{r}_0 - \overline{s}_0) + p)
\right]
\end{equation}
where the cosine terms are replaced by the dot product between the distance vectors and unit vector along the x-axis, $\overline{x}$. The distance between the radar and the first reference RIS element $r_0$ is a known variable. We can multiply the eq. (\ref{eq:y_nVectors}) and get rid of the first term in the complex exponential. This operation is called \textit{focusing} and would yield
\begin{equation}
y_n^{\text{Rx}} = 
\sigma_p A 
\exp \left[-j k_i ( nd\overline{x} \cdot 
(\overline{r}_0 - \overline{s}_0) + s_0  + p) \right]
\end{equation}

Lets assume assume we apply linearly progressing phase shifts in each RIS element given by $\phi_n = k_i n \phi$. This corresponds to far-field beam forming, and it is equivalent to rotating the virtual source around the center of the RIS, which was illustrated in  Fig. \ref{fig:far field}. In this case, the received signal is the combination of all paths over the RIS and becomes,
\begin{equation} \label{eq:y_io}
\begin{split}
y (i,\phi)  &= \sum_{n=0}^{N-1} y_n^{\text{Rx}} = 
\sigma_p A
\sum_{n=0}^{N-1}
e^{-j k_i n d (\phi +   \overline{p} \cdot \overline{x})}
e^{-j k_i (s_0 +p)}
\end{split}
\end{equation}
where the received signal is parametrized by two variables, the fast-time index $i$, and the slow-time index $\phi$. The slow-time index denotes the beamforming angle, which is the direction of the virtual source, and the fast-time index models the propagation of the wavefront. Note that, the summation in the eq. (\ref{eq:y_io}) is in the form of Fourier transform. Defining an auxiliary variable $\Delta = \phi + \overline{p} \cdot \overline{x}$ the expression simplifies as
\begin{equation} 
\begin{split}
y (i,\Delta)  &= 
\frac{1- e^{-j k_i N d \Delta}}{1 - e^{-j k_i d \Delta}}
\sigma_p A
e^{-j k_i (s_0 +p)}
\end{split}
\end{equation} 
Here we observe that the first term is the Fourier transformation of the rectangular window. In source-image perspective, it can be stated that the wavefield generated by the  virtual source  has been \textit{windowed} by the RIS elements, enabling them to propagate in the real environment. To further benefit from this perspective, we might apply different attenuation factors in each RIS element to create different windowing effects, yielding better side-lobe suppression, enhancing azimuth characteristics of  the image. Applying attenuation window at each RIS element by $\Psi_n = w(n) e^{-jk_i\phi n}$ we get the received signal
\begin{equation} 
\begin{split}
y (i,\phi)  &= 
\sigma_p A
\sum_{n=0}^{N-1}
w(n)
e^{-j k_i n d \Delta_p}
e^{-j k_i (s_0 +p)}
\end{split}
\end{equation}
in more compact form,
\begin{equation} \label{eq:y_delt}
\begin{split}
y (i,\Delta_p)  &= 
\sigma_p A
\mathcal{F}_{\Delta_p} \left\{ w  \right \}
e^{-j k_i (s_0 +p)}
\end{split}
\end{equation}
From that, we can obtain the radar image by applying the match filter of eq. (\ref{eq:y_delt}) for each pixel, $y^*(i,\Delta_p)$.